\documentclass[a4paper,11pt]{article}
\pdfoutput=1

\usepackage{prdpub,bm,mathtools,slashed}
\usepackage[usenames,dvipsnames,svgnames,table,x11names]{xcolor}

\usepackage[whole]{bxcjkjatype}
\usepackage[T1]{fontenc}
\usepackage[utf8]{inputenc}

\definecolor{light_blue}{rgb}{0.15, 0.35, 0.95}
\definecolor{kit_green}{rgb}{0, 0.58823, 0.50980}

\usepackage{amsthm}
\usepackage{appendix}
\usepackage{ascmac}
\usepackage{enumerate}
\usepackage{enumitem} 
\usepackage{tensor}
\usepackage{physics}
\usepackage{braket}
\usepackage{mathrsfs}
\usepackage[normalem]{ulem} 
\usepackage{url}
\usepackage[most]{tcolorbox}
\usepackage{indentfirst}
\usepackage{ragged2e}
\usepackage{microtype}
\usepackage{float}
\usepackage{longtable}

\newtcbtheorem[number within=section]{definitionbox}{Definition}{
  enhanced,
  breakable,
  colback=blue!2,
  colframe=blue!40!black,
  boxrule=0.5pt,
  arc=1mm,
  fonttitle=\bfseries,
  left=2mm,
  right=2mm,
  top=1.5mm,
  bottom=1.5mm
}{def}

\newtcolorbox{algorithmbox}[1]{
  enhanced,
  breakable,
  colback=white,
  colframe=black!70,
  boxrule=0.5pt,
  arc=2mm,
  title={#1},
  title after break={#1 (continued)},
  fonttitle=\bfseries,
  coltitle=black,
  colbacktitle=black!6,
  left=2mm,
  right=2mm,
  top=1.5mm,
  bottom=1.5mm,
  before skip=7pt,
  after skip=9pt
}

\usepackage{tikz}
\tikzset{
  qwire/.style={line width=0.45pt},
  qbox/.style={draw, line width=0.55pt, fill=white, align=center},
  hgate/.style={draw, line width=0.55pt, minimum width=0.55cm, minimum height=0.38cm, inner sep=1pt}
}
\newcommand{\wire}[3]{\draw[qwire] (#1,#2) -- (#3,#2);}
\newcommand{\qlabel}[2]{\node[anchor=east] at (-0.35,#1) {$#2$};}
\newcommand{\ctrl}[2]{\fill (#1,#2) circle[radius=1.8pt];}
\newcommand{\targ}[2]{%
  \draw[qwire] (#1,#2) circle[radius=3.5pt];
  \draw[qwire] ({#1-0.09},#2) -- ({#1+0.09},#2);
  \draw[qwire] (#1,{#2-0.09}) -- (#1,{#2+0.09});
}
\newcommand{\cnot}[3]{\draw[qwire] (#1,#2) -- (#1,#3);\ctrl{#1}{#2}\targ{#1}{#3}}
\newcommand{\tof}[4]{\draw[qwire] (#1,#2) -- (#1,#4);\ctrl{#1}{#2}\ctrl{#1}{#3}\targ{#1}{#4}}

\usetikzlibrary{arrows.meta,positioning,shapes.geometric,decorations.pathreplacing}

\makeatletter
\gdef\@fpheader{}
\makeatother

\title{Quantum Algorithms for Finding Vacua in the ADK Landscape}

\author[1]{Shirabe Endo}
\author[1,2,3]{Yuta Hamada}

\affiliation[1]{Graduate Institute for Advanced Studies, SOKENDAI, 1-1 Oho, Tsukuba, Ibaraki 305-0801, Japan}
\affiliation[2]{Theory Center, IPNS, High Energy Accelerator Research Organization (KEK), 1-1 Oho, Tsukuba, Ibaraki 305-0801, Japan}
\affiliation[3]{RIKEN Center for Interdisciplinary Theoretical and Mathematical Sciences (iTHEMS), RIKEN, Wako 351-0198, Japan}

\emailAdd{sendo@post.kek.jp}
\emailAdd{yhamada@post.kek.jp}

\preprint{\shortstack[r]{KEK-TH-2877\\ RIKEN-iTHEMS-Report-26}}

\abstract{
We study quantum algorithms for finding a vacuum with a small cosmological constant in the toy model for string Landscape proposed by Arkani-Hamed, Dimopoulos, and Kachru. 
First, we construct a function-value oracle and use it for Grover search with query complexity $\mathcal{O}(2^{N/2})$, where $N$ is the number of scalar fields. 
Then, we point out that, by exploiting the linearity of the vacuum energy, the problem can be reduced to a collision problem between two blocks of the bit string.
With this structure, we construct two quantum algorithms with query complexity $\mathcal{O}(2^{N/3})$.
Compared with the pseudo-polynomial classical algorithm of Denef and Douglas, the quantum algorithms yield a large reduction in computational cost for the physically motivated parameter values.

}

\begin{document}
\maketitle
\flushbottom

\section{Introduction}
\label{sec:introduction}
The cosmological constant problem is among the most severe fine-tuning problems in fundamental physics.
The observed vacuum energy density, $\Lambda_{\mathrm{obs}}\sim10^{-120}M_{\mathrm{Pl}}^4$~\cite{Planck:2018vyg}, is far smaller than any scale that appears in a theory of quantum gravity.
String theory offers a concrete framework in which such a small value can be accommodated statistically rather than dynamically.
Compactifications with background fluxes~\cite{Giddings:2001yu,Kachru:2003aw} (see Refs.~\cite{Grana:2005jc,Douglas:2006es} for reviews) give rise to a discretuum of vacua labeled by flux quanta. The number of vacua has been estimated as $10^{500}$~\cite{Bousso:2000xa,Susskind:2003kw,Douglas:2003um,Ashok:2003gk,Denef:2004ze}.
The number of such vacua is conjectured to be finite~\cite{Douglas:2005talk,Vafa:2005ui,Acharya:2006zw,Hamada:2021yxy}.
With so many vacua the vacuum energies are finely spaced, and the existence of a vacuum with $\Lambda\sim10^{-120}M_{\mathrm{Pl}}^4$ becomes statistically plausible~\cite{Bousso:2000xa}.
Together with Weinberg's anthropic bound~\cite{Weinberg:1987dv}, this constitutes the landscape approach to the cosmological constant.

The landscape approach raises, however, a question of a different nature. Even if such a vacuum exists, can the universe find it? Denef and Douglas~\cite{Denef:2006ad} pointed out that finding a vacuum whose cosmological constant lies in a prescribed window is an NP-hard problem in the Bousso--Polchinski model~\cite{Bousso:2000xa}, since it contains the subset sum problem as a special case.\footnote{NP-hardness of vacuum and near-vacuum problems has since been established for broader classes of field theories and string compactifications~\cite{Halverson:2018cio}.} 
The same statement holds for the toy landscape of Arkani-Hamed, Dimopoulos, and Kachru (ADK)~\cite{Arkani-Hamed:2005zuc}, in which $N$ scalar fields with two minima each generate $2^N$ vacua whose energies are linear in the binary choice of minima. This model captures the combinatorial structure of the problem, and is the main target of this paper. 
Implications of the computational complexity for cosmology have been discussed in Refs.~\cite{Denef:2017cxt,Bao:2017qmt}.
In particular, Ref.~\cite{Bao:2017qmt} observed that, even though the brute-force cost of scanning $2^N$ vacua exceeds the estimated computational capacity of the observable universe,\footnote{It is formulated as the Computational Censorship Hypothesis.} heuristic algorithms with polynomial average-case cost can nevertheless find vacua with $\Lambda\sim10^{-120}M_{\mathrm{Pl}}^4$ in random ADK landscapes.\footnote{Heuristic and learning-based searches of flux vacua have also been explored~\cite{Halverson:2019tkf,Cole:2019enn}.} These approaches trade worst-case guarantees for average-case efficiency. 
The performance depends on the distribution of the input.

In this paper we take a complementary point of view and ask how far the worst-case cost of the vacuum-search problem can be reduced by quantum computation. More broadly, the interplay between quantum information and quantum gravity has grown into an active field of its own~\cite{Goto:2026mbj}, and the question addressed here is a concrete instance of this interplay on the computational side. It has been noted in the landscape context that Grover search~\cite{Grover:1996qsa} provides a quadratic speedup for unstructured search~\cite{Denef:2006ad}.

Our results are as follows (see also Table~\ref{tab:table1} for an overview of the algorithms). We formulate the ADK vacuum-search problem with an explicit input model and construct a function-value oracle, built from quantum adders, that computes the vacuum energy of a superposition of vacua. Grover search over all $2^N$ vacua, which we refer to as plain Grover search, then finds a vacuum in the target window with query complexity $\mathcal{O}(2^{N/2})$, the quantum counterpart of brute-force search. 
Interestingly, we find that the linearity of the vacuum energy allows a quantum algorithm to do better than plain Grover search.\footnote{Quantum algorithms for vacuum search in the ADK model were also discussed in Ref.~\cite{Bao:2017qmt}, which quoted complexity estimates for existing quantum subset-sum algorithms. Here we focus on explicit oracle constructions, worst-case query complexity, and memory requirements.}
Concretely, we split the bit string labeling a vacuum into two blocks and reduce the problem to a collision problem between the partial sums. This is the structure exploited classically by the meet-in-the-middle algorithm~\cite{Horowitz:1974cpa}, which solves the problem in time $\mathcal{O}^*(2^{N/2})$ with a worst-case guarantee. On the quantum side we construct two algorithms with query complexity $\mathcal{O}(2^{N/3})$, following the quantum algorithms for subset sum~\cite{Bernstein:2013qss}.
One is a Grover search over one block combined with a classically sorted list of the other block (the sorted-list algorithm), and the other is a quantum-walk claw-finding algorithm~\cite{Szegedy:2004um,Tani:2007qb} (the quantum-walk algorithm). 
The first algorithm assumes quantum-accessible classical memory (QRACM)~\cite{Kuperberg:2005dhs,Giovannetti:2008qram}, which we review in App.~\ref{app:qracm}. 
Instead of QRACM, the second algorithm assumes quantum-access quantum memory (QRAQM), which we also review in App.~\ref{app:qraqm}.
Finally, we compare the quantum algorithms with two classical baselines: the pseudo-polynomial algorithm of Denef and Douglas, whose cost grows with $M_{\mathrm{Pl}}^4/\Lambda_{\mathrm{target}}$, and the meet-in-the-middle algorithm, at the physically relevant values $N\sim400$ and $\Lambda_{\mathrm{target}}/M_{\mathrm{Pl}}^4\sim10^{-120}$.

Note that the improvement over the classical algorithm is an improvement of the exponent, which is not a change of the exponential character of the problem.
This is consistent with the expectation that quantum computers do not solve NP-hard problems in polynomial time. 
Our results quantify where the landscape search problem sits with respect to quantum resources.

The paper is organized as follows. In Sec.~\ref{sec:adk-model} we review the ADK model and define the vacuum-search problem. Sec.~\ref{sec:vacuum-search} constructs the algorithms, which are Grover search over all vacua, and the algorithms that exploit the linear structure. Sec.~\ref{sec:classical-quantum-comparison} compares the classical and quantum algorithms. Sec.~\ref{sec:conclusion} summarizes the results and discusses future directions. App.~\ref{sec:grover-search} reviews Grover search, and App.~\ref{app:qram} summarizes the QRAM assumption.

\begin{table}[]
\centering
\begin{tabular}{ccc}
Algorithm                      & Complexity  & Section \\ \cline{1-3}
Grover search                  &     $\mathcal{O}(2^{N/2})$  &~\ref{sec:grover-vacuum-search}      \\
(classical) meet-in-the-middle &    $\mathcal{O}^*(2^{N/2})$  &~\ref{sec:classical-mitm}    \\
sorted-list algorithm &     $\mathcal{O}(2^{N/3})$ &  ~\ref{sec:collision-algorithm} \\
quantum-walk algorithm  &     $\mathcal{O}(2^{N/3})$ & ~\ref{sec:quantum-walk-vacuum-search} \\
(classical) Denef--Douglas & $\mathcal{O}\left(N^3\,M_{\mathrm{Pl}}^4/\Lambda_{\mathrm{target}}\right)$& ~\ref{sec:classical-quantum-comparison}
\end{tabular}
\caption{Summary of algorithms and complexity scalings in this paper. These values are worst-case bounds.}

\label{tab:table1}
\end{table}

\section{ADK model}
\label{sec:adk-model}
We use the toy model proposed by Arkani-Hamed, Dimopoulos, and Kachru~\cite{Arkani-Hamed:2005zuc} (the ADK model). 
In this model, the landscape is mimicked by a field theory with $N$ scalar fields $\phi^i$ and a potential
\begin{align}
    V= \sum^N_{i=1} V_i (\phi^i).
\end{align}
Here each $V_i$ has two local minima $V^a_i,V^b_i$. A vacuum is specified by a choice of which minimum to take for each $V_i$, represented by a parameter $m_i\in\{0,1\}$.
The total number of possible combinations is $2^N$. 
The vacuum energy is then given by
\begin{align}
\Lambda_{\bm m} = \sum^N_{i=1}[ V^a_i+m_i(V^b_i-V^a_i )]=V_0+\sum^N_{i=1}m_i\Delta V_i,
\end{align}
where $V_0=\sum_{i=1}^N V^a_i$ is the vacuum energy when all fields are in the minimum $V^a_i$, and $\Delta V_i=V^b_i-V^a_i$ is the energy difference between the two minima of the $i$-th field.
The problem we consider is to find a vacuum $\bm m\in M$ such that 
\begin{align}
M=\{\bm m\in\{0,1\}^N \mid 0\le\Lambda_{\bm m}\le\Lambda_{\mathrm{target}}\},
\label{eq:marked-vacuum-set}
\end{align}
where $\Lambda_{\mathrm{target}}$ is a given positive constant representing the upper limit of the target window. 
For instance, the realistic value $\Lambda_{\mathrm{target}}\sim10^{-120}M_{\mathrm{Pl}}^4$ is of interest in the context of the cosmological constant problem.
Depending on the choice of $\Lambda_{\mathrm{target}}$, the set $M$ may be empty.

\begin{figure}[htbp]
\centering
\begin{tikzpicture}[
  >=Stealth,
  paneltitle/.style={font=\footnotesize\bfseries, align=center},
  choicebox/.style={draw=blue!55!black, rounded corners=2pt,
    fill=blue!5, thick, align=center, inner sep=7pt},
  flow/.style={-{Stealth[length=3mm]}, thick, draw=black!65}
]

\begin{scope}[shift={(-4.7,0)}]
  \node[paneltitle] at (0,2.35) {(a) Two minima\\for each field};
  \draw[->, semithick] (-1.75,-0.15) -- (1.85,-0.15)
    node[below right=-1pt] {$\phi^i$};
  \draw[->, semithick] (-1.55,-0.40) -- (-1.55,1.75)
    node[above left=-1pt] {$V_i$};
  \draw[very thick, draw=blue!70!black, smooth]
    plot coordinates {
      (-1.45,1.25) (-1.15,0.48) (-0.78,0.05) (-0.38,0.42)
      (0,1.22) (0.38,0.73) (0.78,0.28) (1.15,0.55) (1.55,1.30)
    };
  \fill[OrangeRed] (-0.78,0.05) circle (2.2pt);
  \fill[OrangeRed] (0.78,0.28) circle (2.2pt);
  \node[font=\scriptsize, align=center, below=2pt] at (-0.78,0.05)
    {$V_i^a$\\[-1pt]$(m_i=0)$};
  \node[font=\scriptsize, align=center, below=2pt] at (0.78,0.28)
    {$V_i^b$\\[-1pt]$(m_i=1)$};
  \node[font=\small] at (0,-1.05) {$i=1,\ldots,N$};
\end{scope}

\node[paneltitle] at (0,2.35) {(b) A vacuum
};
\node[choicebox, minimum width=3.45cm, minimum height=1.65cm] (choice) at (0,0.55)
  {$\bm{m}=(m_1,\ldots,m_N)$\\[5pt]
   $\displaystyle \Lambda_{\bm{m}}
     =V_0+\sum_{i=1}^{N}m_i\Delta V_i$};
\node[font=\small] at (0,-0.90) {$2^N$ possible choices};

\begin{scope}[shift={(4.75,0)}]
  \node[paneltitle] at (0,2.35) {(c) Vacuum energy};
  \fill[green!12] (-0.85,-0.72) rectangle (0.85,-0.34);
  \draw[green!55!black, thin] (-0.85,-0.72) rectangle (0.85,-0.34);
  \draw[->, semithick] (0,-1.22) -- (0,1.75)
    node[above] {$\Lambda$};
  \draw[thin] (-0.16,-0.72) -- (0.16,-0.72);
  \node[font=\scriptsize, left=2pt] at (-0.16,-0.72) {$0$};
  \node[font=\scriptsize, text=green!45!black, right=3pt, align=left]
    at (0.85,-0.53) {target\\window};

  \foreach \y in {-1.02,-0.55,-0.10,0.27,0.63,1.02,1.38} {
    \draw[black!55] (-0.32,\y) -- (0.32,\y);
  }
  \fill[black!65] (0,-1.02) circle (1.5pt);
  \fill[green!55!black] (0,-0.55) circle (2.3pt);
  \fill[black!65] (0,-0.10) circle (1.5pt);
  \fill[black!65] (0,0.27) circle (1.5pt);
  \fill[black!65] (0,0.63) circle (1.5pt);
  \fill[black!65] (0,1.02) circle (1.5pt);
  \fill[black!65] (0,1.38) circle (1.5pt);
  \node[font=\small, align=center] at (1.25,0.65) {$\vdots$\\$2^N$ vacua};
\end{scope}

\draw[flow] (-2.75,0.65) -- (choice.west);
\draw[flow] (choice.east) -- (3.15,0.65);

\end{tikzpicture}
\caption{Schematic illustration of the ADK landscape. Each field $\phi^i$ has two local minima, encoded by a binary variable $m_i$. A bit string $\bm{m}$ specifies one of the $2^N$ vacua, whose energy is obtained by summing the selected contributions. The shaded band denotes the target range for the cosmological constant.}
\label{fig:adk-landscape-schematic}
\end{figure}
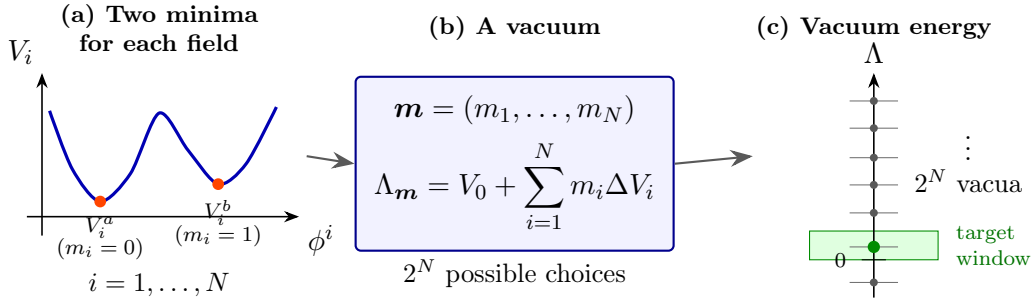

In Fig.~\ref{fig:adk-landscape-schematic}, we schematically illustrate the ADK landscape, showing how the vacuum energy is determined by the choice of local minima for each potential.

The vacuum-search problem in the ADK model is summarized as follows.
\begin{definitionbox}{ADK vacuum-search problem}{adk-vacuum-search-problem}
The input of the ADK vacuum-search problem consists of an integer $N$, the values $V_0$ and $\Delta V_i\ (i=1,2,\ldots,N)$, and a positive constant $\Lambda_{\mathrm{target}}$ representing the upper limit of the window $[0,\Lambda_{\mathrm{target}}]$. 
The output is a bit string
\begin{align}
\bm{m}=(m_1,m_2,\ldots,m_N)\in\{0,1\}^N,
\label{eq:adk-vacuum-search-output}
\end{align}
which belongs to the set $M$ defined in Eq.~\eqref{eq:marked-vacuum-set}, i.e., an element of the set of vacua satisfying the target window condition.
If $M=\varnothing$, the output is ``no solution''.
The quantum algorithms should output an element of $M$ with
probability at least $2/3$ whenever $M\neq\varnothing$.

\end{definitionbox}
Note that the computational complexity of the vacuum search in the ADK model has been discussed in Ref.~\cite{Bao:2017qmt}, including heuristic algorithms, as mentioned in Sec.~\ref{sec:introduction}. In this paper, we aim to reduce the computational complexity with worst-case guarantees using quantum algorithms.

\section{Vacuum-search algorithms for the ADK model}
\label{sec:vacuum-search}
In this section, we construct classical and quantum algorithms for the ADK vacuum-search problem defined in Definition~\ref{def:adk-vacuum-search-problem}, and evaluate their computational complexity and required resources. 
We use $\mathcal{O}^*$ notation to suppress factors polynomial in $N$ and in the bit length of the encoded energy parameters.
Our complexity comparisons focus on the remaining exponential dependence.
Bounds on explicitly specified query counts are stated using ordinary $\mathcal{O}$ notation.
In the main text, query complexity counts calls to the function-value oracles.
QRACM accesses are counted separately.
In Sec.~\ref{sec:vacuum-oracle}, we explain how to encode $\bm m$ and $\Lambda_{\bm m}$ in qubits, and construct a function-value oracle $U_\Lambda$ that computes $\Lambda_{\bm m}$ from $\bm m$ using a quantum adder. 
This oracle is common to all subsequent algorithms. In Sec.~\ref{sec:grover-vacuum-search}, we consider the most basic algorithm that applies Grover search~\cite{Grover:1996qsa} to all $2^N$ vacua (hereafter referred to as plain Grover search), and show that it has a query complexity of $\mathcal{O}(2^{N/2})$. In Sec.~\ref{sec:structured-vacuum-search}, we exploit the structure of the vacuum energy being a linear sum to realize $\mathcal{O}(2^{N/3})$ with two types of algorithms (the sorted-list algorithm and the quantum-walk algorithm).

\subsection{Qubit register and function-value oracle for vacuum energy}
\label{sec:vacuum-oracle}
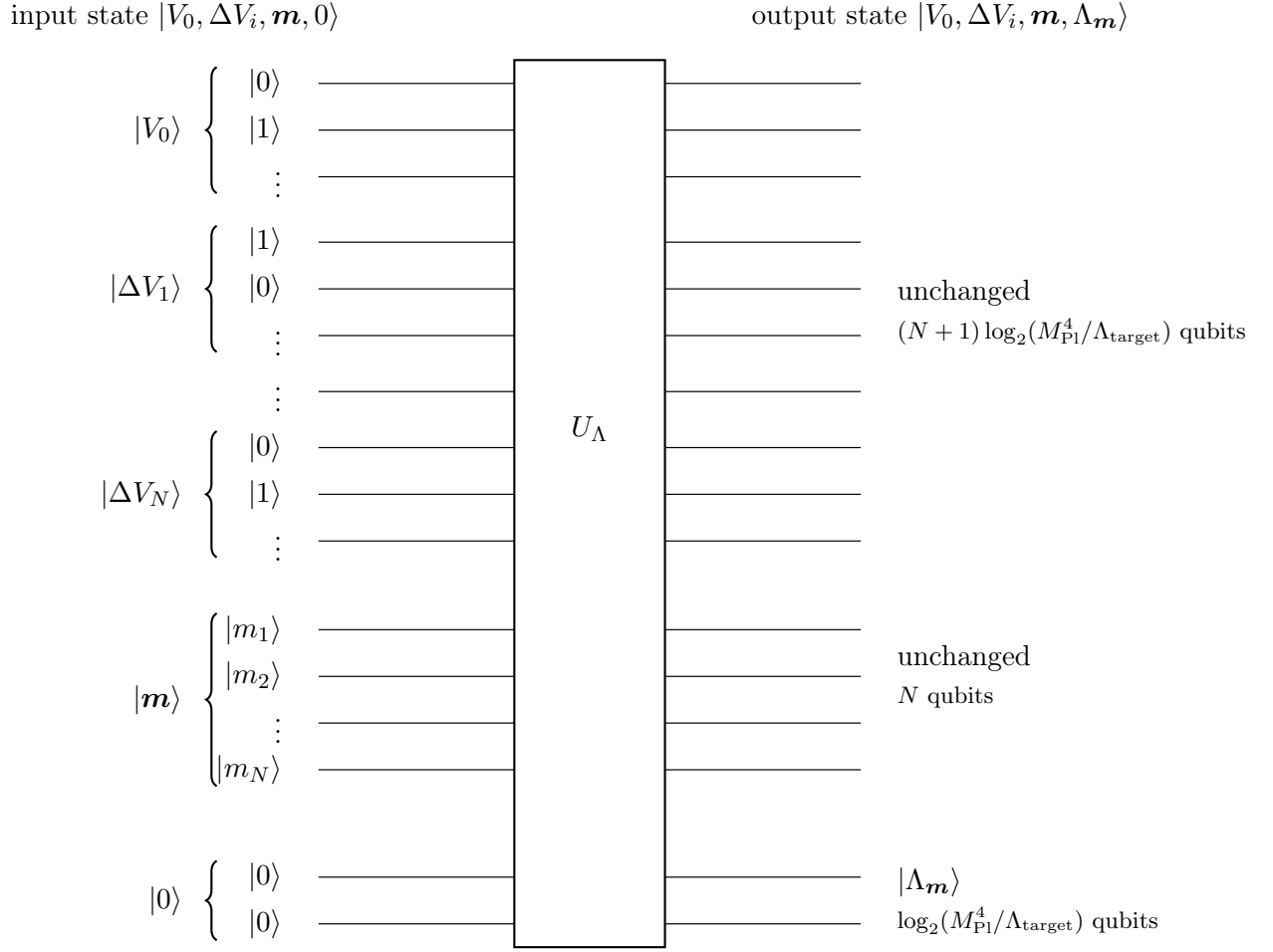
\begin{figure}[htbp]
\centering
\begin{tikzpicture}[x=1cm,y=0.62cm,
  brace/.style={decorate,decoration={brace,amplitude=4pt},thick},
  regnote/.style={font=\footnotesize,anchor=west}
]
  \def\yA{0}     \def\yB{-1}      \def\yC{-2}            
  \def\yD{-3.4}  \def\yE{-4.4}     \def\yF{-5.4}         
  \def\yG{-6.6}                                    
  \def\yH{-7.8}  \def\yI{-8.8}     \def\yJ{-9.8}        
  \def\yK{-11.7}  \def\yL{-12.7} \def\yM{-13.7} \def\yN{-14.7}   
  \def\yO{-17.0} \def\yP{-18.0}                   
  \foreach \y in {\yA,\yB,\yC,\yD,\yE,\yF,\yG,\yH,\yI,\yJ,\yK,\yL,\yM,\yN,\yO,\yP}{\wire{0}{\y}{7.2}}
  \node[anchor=east] at (-0.35,\yA) {$\ket{0}$};
  \node[anchor=east] at (-0.35,\yB) {$\ket{1}$};
  \node[anchor=east] at (-0.35,\yC) {$\vdots$};
  \node[anchor=east] at (-0.35,\yD) {$\ket{1}$};
  \node[anchor=east] at (-0.35,\yE) {$\ket{0}$};
  \node[anchor=east] at (-0.35,\yF) {$\vdots$};
  \node[anchor=east] at (-0.35,\yG) {$\vdots$};
  \node[anchor=east] at (-0.35,\yH) {$\ket{0}$};
  \node[anchor=east] at (-0.35,\yI) {$\ket{1}$};
  \node[anchor=east] at (-0.35,\yJ) {$\vdots$};
  \node[anchor=east] at (-0.35,\yK) {$\ket{m_1}$};
  \node[anchor=east] at (-0.35,\yL) {$\ket{m_2}$};
  \node[anchor=east] at (-0.35,\yM) {$\vdots$};
  \node[anchor=east] at (-0.35,\yN) {$\ket{m_N}$};
  \node[anchor=east] at (-0.35,\yO) {$\ket{0}$};
  \node[anchor=east] at (-0.35,\yP) {$\ket{0}$};
  \node at (3.6,\yC) {$\vdots$};
  \node at (3.6,\yF) {$\vdots$};
  \node at (3.6,\yG) {$\vdots$};
  \node at (3.6,\yJ) {$\vdots$};
  \node at (3.6,\yM) {$\vdots$};
  \node[anchor=south] at (-1.9,{\yA+0.9}) {input state $\ket{V_0,\Delta V_i,\bm m,0}$};
  \node[anchor=south east] at (10.9,{\yA+0.9}) {output state $\ket{V_0,\Delta V_i,\bm m,\Lambda_{\bm m}}$};
  \draw[brace] (-1.35,{\yC-0.35}) -- (-1.35,{\yA+0.35}) node[midway,xshift=-9pt,anchor=east] {$\ket{V_0}$};
  \draw[brace] (-1.35,{\yF-0.35}) -- (-1.35,{\yD+0.35}) node[midway,xshift=-9pt,anchor=east] {$\ket{\Delta V_1}$};
  \draw[brace] (-1.35,{\yJ-0.35}) -- (-1.35,{\yH+0.35}) node[midway,xshift=-9pt,anchor=east] {$\ket{\Delta V_N}$};
  \draw[brace] (-1.35,{\yN-0.35}) -- (-1.35,{\yK+0.35}) node[midway,xshift=-9pt,anchor=east] {$\ket{\bm m}$};
  \draw[brace] (-1.35,{\yP-0.35}) -- (-1.35,{\yO+0.35}) node[midway,xshift=-9pt,anchor=east] {$\ket{0}$};
  \draw[thick,fill=white] (2.6,{\yP-0.5}) rectangle (4.6,{\yA+0.5});
  \node at (3.6,{(\yA+\yP)/2+1.6}) {$U_\Lambda$};
  \node[anchor=west,align=left] at (7.55,{(\yA+\yJ)/2}) {unchanged\\[1pt]{\footnotesize $(N+1)\log_2(M_{\mathrm{Pl}}^4/\Lambda_{\mathrm{target}})$ qubits}};
  \node[anchor=west,align=left] at (7.55,{(\yK+\yM)/2}) {unchanged\\[1pt]{\footnotesize $N$ qubits}};
  \node[anchor=west,align=left] at (7.55,{(\yO+\yP)/2}) {$\ket{\Lambda_{\bm m}}$\\[1pt]{\footnotesize $\log_2(M_{\mathrm{Pl}}^4/\Lambda_{\mathrm{target}})$ qubits}};
\end{tikzpicture}
\caption{
The configuration of the registers representing the state $\ket{V_0,\Delta V_i,\bm m,0}$ on which the oracle $U_\Lambda$ acts.  
}
\label{fig:register-layout}
\end{figure}

We first embed the information necessary for the computation into qubits. 
We denote the state by $\ket{V_0,\Delta V_i,\bm m,0}$, which consists of four types of registers.
The first register represents $V_0$, and the second register represents $\Delta V_1,\ldots,\Delta V_N$, each encoded in binary as a computational basis state.\footnote{Throughout, the energy parameters are treated as exact finite binary inputs. The window condition in Eq.~\eqref{eq:marked-vacuum-set} is imposed on these encoded values. All intermediate sums are represented with enough bits to avoid overflow.} 
We take the number of qubits used for each input energy value to be approximately $\log_2(M_{\mathrm{Pl}}^4/\Lambda_{\mathrm{target}})$.
These registers are provided as inputs and remain unchanged during the execution of the algorithm. 
The third register consists of $N$ qubits representing $\bm m$, where the $i$-th qubit holds $m_i\in\{0,1\}$ in the computational basis.
In the subsequent Grover search, a superposition of all $2^N$ possible $\bm m$ values is placed in this register.
The fourth register is a value register for writing the computation result $\Lambda_{\bm m}$.
Negative values are handled using two's complement representation. The initial state of this register is $\ket{0}$.
The configuration is illustrated in Fig.~\ref{fig:register-layout}.

\begin{figure}[htbp]
\centering
\begin{tikzpicture}[
  x=1cm,
  y=0.75cm,
  wire/.style={thick},
  control/.style={
    circle,
    fill=black,
    inner sep=0pt,
    minimum size=4.5pt
  },
  target/.style={
    circle,
    draw=black,
    thick,
    inner sep=0pt,
    minimum size=10pt
  },
  adder/.style={
    draw=black,
    very thick,
    fill=white
  },
  every node/.style={font=\large}
]

\def\xL{3.0}
\def\xR{13.25}

\foreach \y in {
  0,-1,
  -2.3,-3.3,
  -4.6,-5.6,
  -6.9,-7.9,
  -9.2,-10.2,-11.2,-12.2,-13.2,
  -14.2,-15.2,-16.2,-17.2
}{
  \draw[wire] (\xL,\y) -- (\xR,\y);
}

\node[anchor=east] at (0.55,-0.5) {$\ket{V_0}$};

\node[anchor=east] at (2.60,0)  {$\ket{0}$};
\node[anchor=east] at (2.60,-1) {$\ket{1}$};

\draw[thick]
  (1.65,0.58)
  .. controls (1.20,0.42) and (1.20,-1.42) ..
  (1.65,-1.58);

\node[anchor=east] at (0.55,-2.8)
  {$\ket{\Delta V_1}$};

\node[anchor=east] at (2.60,-2.3) {$\ket{1}$};
\node[anchor=east] at (2.60,-3.3) {$\ket{0}$};

\draw[thick]
  (1.65,-1.72)
  .. controls (1.20,-1.88) and (1.20,-3.72) ..
  (1.65,-3.88);

\node[anchor=east] at (0.55,-5.1)
  {$\ket{\Delta V_2}$};

\node[anchor=east] at (2.60,-4.6) {$\ket{1}$};
\node[anchor=east] at (2.60,-5.6) {$\ket{1}$};

\draw[thick]
  (1.65,-4.02)
  .. controls (1.20,-4.18) and (1.20,-6.02) ..
  (1.65,-6.18);

\node[anchor=east] at (2.60,-6.9) {$\ket{m_1}$};
\node[anchor=east] at (2.60,-7.9) {$\ket{m_2}$};

\foreach \y in {
  -9.2,-10.2,-11.2,-12.2,-13.2,
  -14.2,-15.2,-16.2,-17.2
}{
  \node[anchor=east] at (2.60,\y) {$\ket{0}$};
}

\draw[wire] (3.55,0) -- (3.55,-9.2);
\node[control] at (3.55,0) {};
\node[target] at (3.55,-9.2) {};

\draw[thick]
  (3.41,-9.2) -- (3.69,-9.2);
\draw[thick]
  (3.55,-9.34) -- (3.55,-9.06);

\draw[wire] (4.20,-1) -- (4.20,-10.2);
\node[control] at (4.20,-1) {};
\node[target] at (4.20,-10.2) {};

\draw[thick]
  (4.06,-10.2) -- (4.34,-10.2);
\draw[thick]
  (4.20,-10.34) -- (4.20,-10.06);

\draw[wire] (4.95,-2.3) -- (4.95,-11.2);

\node[control] at (4.95,-2.3) {};
\node[control] at (4.95,-6.9) {};
\node[target] at (4.95,-11.2) {};

\draw[thick]
  (4.81,-11.2) -- (5.09,-11.2);
\draw[thick]
  (4.95,-11.34) -- (4.95,-11.06);

\draw[wire] (5.65,-3.3) -- (5.65,-12.2);

\node[control] at (5.65,-3.3) {};
\node[control] at (5.65,-6.9) {};
\node[target] at (5.65,-12.2) {};

\draw[thick]
  (5.51,-12.2) -- (5.79,-12.2);
\draw[thick]
  (5.65,-12.34) -- (5.65,-12.06);

\draw[wire] (6.40,-4.6) -- (6.40,-14.2);

\node[control] at (6.40,-4.6) {};
\node[control] at (6.40,-7.9) {};
\node[target] at (6.40,-14.2) {};

\draw[thick]
  (6.26,-14.2) -- (6.54,-14.2);
\draw[thick]
  (6.40,-14.34) -- (6.40,-14.06);

\draw[wire] (7.10,-5.6) -- (7.10,-15.2);

\node[control] at (7.10,-5.6) {};
\node[control] at (7.10,-7.9) {};
\node[target] at (7.10,-15.2) {};

\draw[thick]
  (6.96,-15.2) -- (7.24,-15.2);
\draw[thick]
  (7.10,-15.34) -- (7.10,-15.06);

\node[
  adder,
  minimum width=2.10cm,
  minimum height=3.45cm
]
at (8.65,-11.2)
{Adder};

\node[
  adder,
  minimum width=2.15cm,
  minimum height=5.05cm
]
at (11.15,-14.2)
{Adder};

\draw[thick]
  (13.45,-13.55)
  .. controls
     (14.05,-13.75)
     and
     (14.05,-17.65)
  ..
  (13.45,-17.85);

\node[anchor=west]
  at (14.55,-15.70)
  {$\ket{\Lambda_{\bm m}}$};

\end{tikzpicture}
\caption{Example of the function-value oracle in Eq.~\eqref{eq:vacuum-energy-oracle}, implemented using quantum adders (shown as {\it Adder} in the figure).
The inverse operations that reset the auxiliary registers to
$\ket{0}$ are omitted from the figure.
}
\label{fig:vacuum-energy-oracle-circuit}
\end{figure}
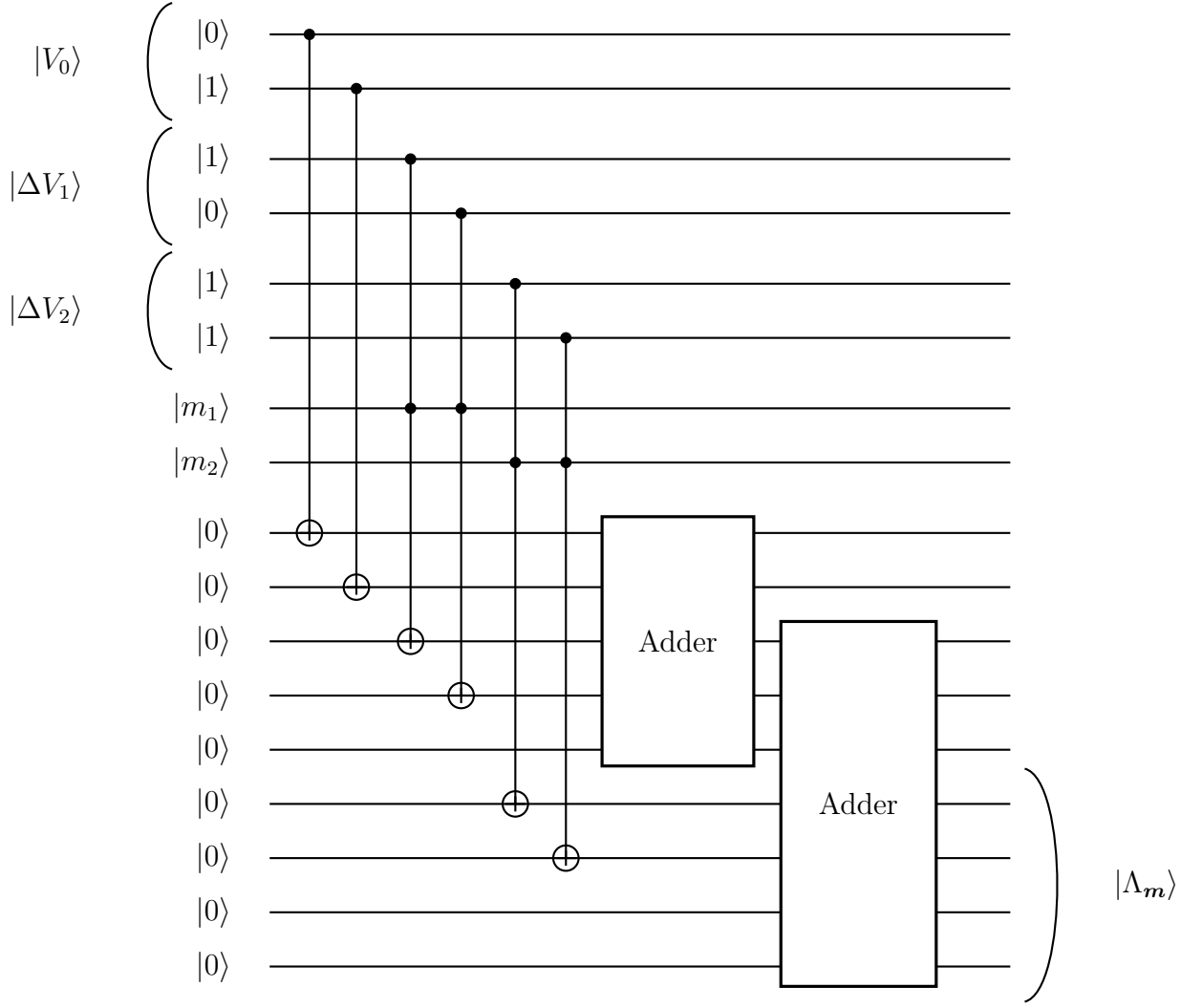

Then, we define a function-value oracle $U_\Lambda$ that acts on the state corresponding to $\bm m$ as follows:
\begin{align}
U_\Lambda\ket{V_0,\Delta V_i,\bm m,0} = \ket{V_0,\Delta V_i,\bm m,\Lambda_{\bm m}}.
\label{eq:vacuum-energy-oracle}
\end{align}
All the algorithms in this section  use this oracle.
This oracle is constructed from quantum adders that add the binary inputs $V_0$ and $\Delta V_i$, controlled by the register representing $\bm m$. 
An example of a quantum adder is shown in Fig.~\ref{fig:first-circuit},\footnote{
There are various types of quantum adders that have been proposed in recent years, which are optimized from the perspective of quantum resources by utilizing techniques such as quantum Fourier transforms~\cite{Wang:2024qac}.} where the logic circuit appearing in the classical adder is replaced by quantum gates~\cite{Vedral:1995qf}.
An example of the function-value oracle built from such adders is shown in Fig.~\ref{fig:vacuum-energy-oracle-circuit}.

\begin{figure}[htbp]
\centering
\begin{tikzpicture}[x=0.86cm,y=0.64cm]
  \foreach \y in {0,-1,-2,-3,-4.35,-5.35,-6.35}{\wire{0}{\y}{10.4}}
  \qlabel{0}{\ket{0}}
  \qlabel{-1}{\ket{a_1}}
  \qlabel{-2}{\ket{b_1}}
  \qlabel{-3}{\ket{0}}
  \qlabel{-4.35}{\ket{a_2}}
  \qlabel{-5.35}{\ket{b_2}}
  \qlabel{-6.35}{\ket{0}}

  \tof{0.75}{-1}{-2}{-3}
  \cnot{1.35}{-1}{-2}
  \tof{2.10}{0}{-2}{-3}

  \tof{3.00}{-4.35}{-5.35}{-6.35}
  \cnot{3.55}{-4.35}{-5.35}
  \tof{4.15}{-3}{-5.35}{-6.35}
  \cnot{4.70}{-4.35}{-5.35}
  \cnot{5.25}{-4.35}{-5.35}
  \cnot{5.85}{-3}{-5.35}

  \tof{6.75}{0}{-2}{-3}
  \cnot{7.30}{-1}{-2}
  \tof{7.85}{-1}{-2}{-3}
  \cnot{8.42}{-1}{-2}
  \cnot{9.05}{0}{-2}
\end{tikzpicture}
\caption{
An example of a quantum adder, where the addition of two binary numbers $a_2a_1$ and $b_2b_1$ is performed. The circuit acts as $\ket{a,b}\rightarrow\ket{a,b+a}$.  
}
\label{fig:first-circuit}
\end{figure}
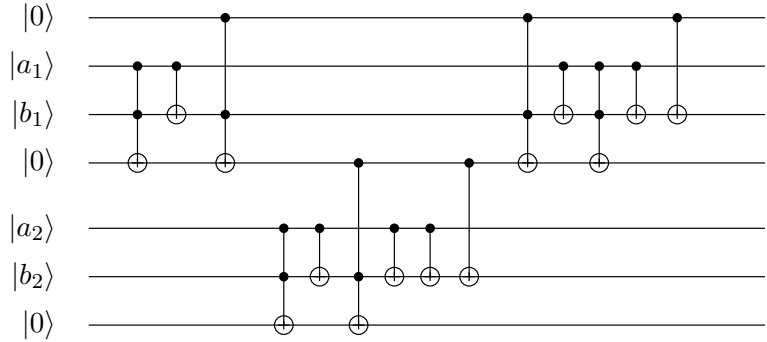

\subsection{Quantum vacuum search by Grover search}
\label{sec:grover-vacuum-search}
As a first quantum algorithm for the ADK vacuum-search problem, we apply Grover search (see App.~\ref{sec:grover-search} for a review). We refer to this algorithm as plain Grover search. 
The initial state $\ket{\phi_0}$ is chosen as the equal superposition of all $2^N$ states:
\begin{align}
\ket{\phi_0}= \frac{1}{\sqrt{2^N}} \sum_{\bm m} \ket{V_0,\Delta V_i,\bm m,0}.
\end{align}
As reviewed in App.~\ref{sec:grover-search}, we need a phase oracle that flips the sign of the states corresponding to the solutions. In our case, the solutions are the states $\ket{V_0,\Delta V_i,\bm m,0}$ for which $0\le\Lambda_{\bm m}\le\Lambda_{\mathrm{target}}$. Therefore, we define the operator $U_M$ that flips the sign of the states corresponding to the solutions:
\begin{align}
U_M\ket{V_0,\Delta V_i,\bm m,\Lambda_{\bm m}} = 
\begin{cases}
\ket{V_0,\Delta V_i,\bm m,\Lambda_{\bm m}} & \text{if} \ \bm m \notin M\\
-\ket{V_0,\Delta V_i,\bm m,\Lambda_{\bm m}}                 &  \text{if} \ \bm m \in M 
\end{cases}.
\label{eq:vacuum-phase-flip}
\end{align}
This operator can be implemented as a controlled operation that checks whether the sign bit of the value register is $0$ (i.e., $\Lambda_{\bm m}\ge0$) and whether $\Lambda_{\bm m}\le\Lambda_{\mathrm{target}}$ using a comparator, and flips the phase only when both conditions are satisfied. The comparator can be constructed from a reversible classical logic circuit, similar to the quantum adder.
Then, the phase oracle $U_{\mathrm{Grover}}$ is constructed as
\begin{align}
U_{\mathrm{Grover}}=U_\Lambda^\dagger U_M U_\Lambda,\qquad
U_{\mathrm{Grover}}\ket{V_0,\Delta V_i,\bm m,0}=
\begin{cases}
\ket{V_0,\Delta V_i,\bm m,0} & \text{if} \ \bm m \notin M\\
-\ket{V_0,\Delta V_i,\bm m,0} & \text{if} \ \bm m \in M
\end{cases}.
\label{eq:vacuum-grover-oracle}
\end{align}

By using $U_{\mathrm{Grover}}$ and the diffusion operator $2\ket{\phi_0}\bra{\phi_0}-I$ with the initial state $\ket{\phi_0}$, we perform Grover search as described in App.~\ref{sec:grover-search}. The query complexity is
\begin{align}
\mathcal{O}(2^{N/2}).
\end{align}

\subsection{Further speedup using the structure of the problem}
\label{sec:structured-vacuum-search}
In the ADK model, we can do better than the plain Grover search by utilizing the fact that $\Lambda_{\bm m}$ is a linear function of the binary parameters $\bm m$. 
In this subsection, we will utilize this structure to further reduce the computational complexity. 
The key is that if we divide the bit string into two blocks, the vacuum energy can be decomposed into the sum of contributions from the two blocks.
This allows the vacuum-search problem to be reduced to a collision problem, where we look for pairs of contributions that sum to a value within the specified range (Sec.~\ref{sec:collision-translation}).
The classical algorithm known as meet-in-the-middle utilizes this structure, providing a speedup from $\mathcal{O}^*(2^N)$ to $\mathcal{O}^*(2^{N/2})$ (Sec.~\ref{sec:classical-mitm}).
In this subsection, we will apply two quantum algorithms for the collision problem to the vacuum-search problem for further speedup.
One algorithm keeps one block of the contributions as a sorted list and performs Grover search on the other block (referred to as the sorted-list algorithm), achieving a query complexity of $\mathcal{O}(2^{N/3})$ (Sec.~\ref{sec:collision-algorithm}).
The other algorithm uses a quantum walk for claw finding (referred to as the quantum-walk algorithm), also achieving a complexity of $\mathcal{O}(2^{N/3})$ (Sec.~\ref{sec:quantum-walk-vacuum-search}).
We will refer to both algorithms collectively as collision-based algorithms.
While these two algorithms may seem different, they lead to the same computational complexity results.\footnote{However, the types of resources required are different.
The sorted-list algorithm requires $\mathcal{O}(N+\log_2(M_{\mathrm{Pl}}^4/\Lambda_{\mathrm{target}}))$ qubits and exponentially large classical memory under QRACM (App.~\ref{app:qracm}), while the quantum-walk algorithm requires the same amount as a quantum register.}

\subsubsection{Reduction to a collision problem}
\label{sec:collision-translation}
The ADK vacuum-search problem can be reduced to a collision problem by dividing the $N$ variables into two parts:
\begin{align}
&\bm m=(x,y),
&&x=(m_1,\ldots,m_p)\in X\equiv\{0,1\}^p,
&&y=(m_{p+1},\ldots,m_N)\in Y\equiv\{0,1\}^{N-p}.
\end{align}
The vacuum energy is expressed as
\begin{align}
&\Lambda_{\bm m}=\Lambda_X(x)+\Lambda_Y(y),
&&\Lambda_X(x)=V_0+\sum^{p}_{i=1}m_i\Delta V_i,
&&\Lambda_Y(y)=\sum^{N}_{i=p+1}m_i\Delta V_i.
\end{align}
Note that the numbers of elements in the sets $X$ and $Y$ are $|X|=2^p$ and $|Y|=2^{N-p}$, respectively.
Now the vacuum-search problem is to find a pair $(x,y)\in X\times Y$ such that
\begin{align}
0 \le\Lambda_X(x)+\Lambda_Y(y) \le \Lambda_{\mathrm{target}}.
\end{align}
This type of problem, where we look for pairs of values from two functions that satisfy a specified relation, is referred to as a collision problem in this paper.
\subsubsection{Classical algorithm: meet-in-the-middle}
\label{sec:classical-mitm}
\begin{figure*}[htbp]
\centering
\begin{tikzpicture}[x=1cm,y=0.5cm,
  cell/.style={draw=black!70,minimum width=1.5cm,minimum height=0.46cm,inner sep=1pt,font=\scriptsize,anchor=north,fill=white},
  hit/.style={fill=green!20},
  lab/.style={font=\scriptsize},
  head/.style={font=\footnotesize\bfseries,anchor=south}
]
\node[head] at (1.6,3.4) {(a) Building the sorted list $L_X$};
\node[lab,anchor=south,align=center] at (0,0.15) {enumerate $x\in X$\\$(x,\Lambda_X(x))$};
\foreach \i/\x/\v in {0/000/2,1/001/-6,2/010/6,3/011/-2,4/100/0,5/101/-8,6/110/4,7/111/-4} {
  \node[cell] at (0,-\i) {$(\x,\ \v)$};
}
\draw[-{Stealth[length=2.5mm]},thick] (0.9,-4.0) -- (2.3,-4.0)
  node[midway,above=1pt,lab,align=center] {sort by\\$\Lambda_X(x)$};
\node[lab,anchor=south,align=center] at (3.2,0.15) {$L_X$\\ascending in $\Lambda_X(x)$};
\foreach \i/\x/\v in {0/101/-8,1/001/-6,2/111/-4,3/011/-2,4/100/0,5/000/2,6/110/4,7/010/6} {
  \node[cell] at (3.2,-\i) {$(\x,\ \v)$};
}
\begin{scope}[shift={(7.4,0)}]
\node[head] at (1.3,3.4) {(b) Binary search in $L_X$ for a given $y$};
\node[lab,anchor=south,align=center] at (1.3,1.7) {given $y$ with $\Lambda_Y(y)=5$ and $\Lambda_{\mathrm{target}}=2$:\\target interval $[-\Lambda_Y(y),\,\Lambda_{\mathrm{target}}-\Lambda_Y(y)]=[-5,-3]$};
\node[lab,anchor=south,align=center] at (0,0.15) {$L_X$\\$(x,\Lambda_X(x))$};
\foreach \i/\x/\v in {0/101/-8,1/001/-6,2/111/-4,3/011/-2,4/100/0,5/000/2,6/110/4,7/010/6} {
  \node[cell] at (0,-\i) {$(\x,\ \v)$};
}
\node[cell,hit] at (0,-2) {$(111,\ -4)$};
\draw[decorate,decoration={brace,amplitude=3pt,mirror},thick,green!50!black]
  (-0.95,-2.92) -- (-0.95,-2.0)
  node[midway,left=4pt,lab,align=right,text=green!40!black] {target\\interval\\$[-5,-3]$};
\draw[thick,black!60] (1.0,0) -- (1.0,-7.92);
\draw[thick,black!60] (0.92,0) -- (1.08,0);  \draw[thick,black!60] (0.92,-7.92) -- (1.08,-7.92);
\fill[black!80] (1.0,-3.46) circle (2.2pt);
\draw[thick,black!60] (1.4,0) -- (1.4,-2.92);
\draw[thick,black!60] (1.32,0) -- (1.48,0);  \draw[thick,black!60] (1.32,-2.92) -- (1.48,-2.92);
\fill[black!80] (1.4,-1.46) circle (2.2pt);
\draw[thick,green!50!black] (1.8,-2.0) -- (1.8,-2.92);
\fill[green!50!black] (1.8,-2.46) circle (2.2pt);
\node[lab,anchor=west] at (2.0,-3.46) {iteration 1: $-2$ too large};
\node[lab,anchor=west] at (2.0,-1.46) {iteration 2: $-6$ too small};
\node[lab,anchor=west,text=green!40!black] at (2.0,-2.46) {iteration 3: $-4$ found};
\end{scope}
\end{tikzpicture}
\caption{
Schematic illustration of list construction and binary search ($p=3$ example with $V_0=2$ and $(\Delta V_1,\Delta V_2,\Delta V_3)=(-2,4,-8)$). Each cell represents a pair $(x,\Lambda_X(x))$. (a) Compute $\Lambda_X(x)$ for all elements $x\in X$ and create a list $L_X$ sorted in ascending order of $\Lambda_X(x)$. (b) For a given $y$, perform a binary search\protect\footnotemark{} in $L_X$ to find $x$ such that $\Lambda_X(x)$ is within the interval $[-\Lambda_Y(y),\Lambda_{\mathrm{target}}-\Lambda_Y(y)]$. The figure shows the case of $\Lambda_Y(y)=5$ and $\Lambda_{\mathrm{target}}=2$. The right vertical lines indicate the search range in each iteration of the binary search, and the black dots represent the middle elements. After three iterations, the element $x=111$ (green) that falls within the target interval is found.}
\label{fig:mitm-schematic}
\end{figure*}
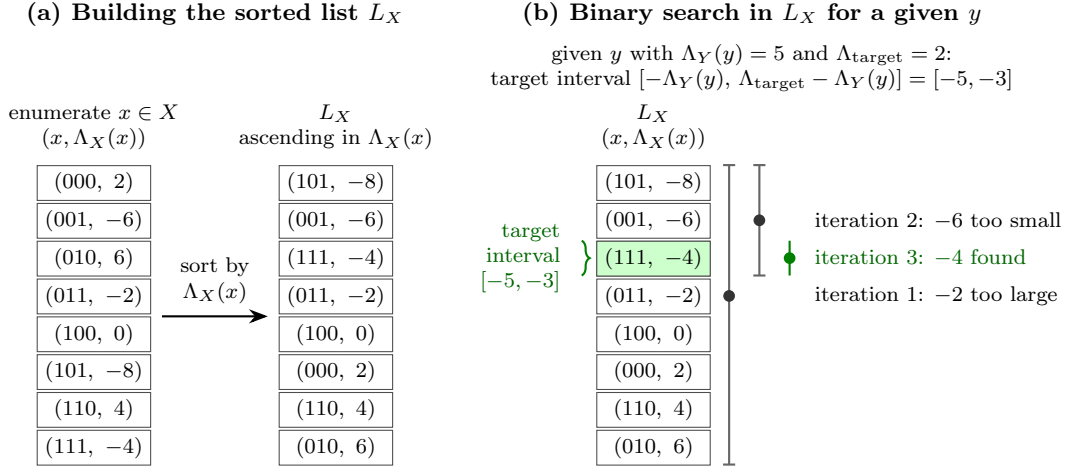

Utilizing the structure of the collision problem, we can apply the classical algorithm known as meet-in-the-middle~\cite{Horowitz:1974cpa} to the vacuum-search problem, which is faster than a brute-force search over all $2^N$ vacua. In the following, we set $p=N/2$.
The algorithm consists of the following steps:
\begin{enumerate}
    \item Compute $\Lambda_X(x)$ for all elements $x\in X$ and create a list $L_X$ of pairs $(x,\Lambda_X(x))$ sorted in ascending order of $\Lambda_X(x)$ (Fig.~\ref{fig:mitm-schematic}(a)). 
    \item Choose an element $y\in Y$ and search the list $L_X$ by binary search for an element $x$ satisfying $0\leq \Lambda_X(x)+\Lambda_Y(y)\leq \Lambda_{\mathrm{target}}$ (see Fig.~\ref{fig:mitm-schematic}(b)). If no such element exists, choose another element of $Y$ and repeat the same binary-search procedure.

    \item Terminate the search when the sum satisfies $0\le\Lambda_X(x)+\Lambda_Y(y)\le\Lambda_{\mathrm{target}}$, and output $\bm m=(x,y)$. If all the elements of $Y$ have been examined, output that no solution exists.
\end{enumerate}

\footnotetext{The binary search is a procedure for finding an element corresponding to a specified value in a sorted list. We first compare the specified value with the middle element of the list. If the specified value is smaller, we restrict the candidates to the first half while if it is larger, we restrict the candidates to the second half. We repeat the same operation iteratively. Since the candidates are halved with each comparison, we can find the corresponding element in a list of $K$ elements with $\lceil\log_2 K\rceil$ comparisons.}

Let us evaluate the computational complexity of the algorithm.
In step 1, generating the list requires $|X|=2^{N/2}$ partial sum calculations and $\mathcal{O}(2^{N/2}\log 2^{N/2})=\mathcal{O}(N2^{N/2})$ comparisons for sorting.
The scan in step 2 requires only $\mathcal{O}(2^{N/2})$ computations. Therefore, suppressing polynomial factors, this classical algorithm can perform vacuum search with a computational complexity of
\begin{align}
\mathcal{O}^*(2^{N/2}).
\end{align}
Compared to the brute-force search of $\mathcal{O}^*(2^N)$, the exponent is halved, which is due to dividing the variables into two blocks, enumerating the partial sums independently, and matching them in linear time.

On the other hand, this algorithm requires storing the list in memory. The list has $2^{N/2}$ elements, and each element consists of $N/2$ bits representing $x$ and $\mathcal{O}(\log_2(M_{\mathrm{Pl}}^4/\Lambda_{\mathrm{target}}))$ bits representing the contribution to the vacuum energy. Therefore, to use this algorithm, we need\footnote{Note that the memory requirement can be reduced from $2^{N/2}$ to $2^{N/4}$ by optimizing the way of splitting~\cite{Schroeppel:1981np}.}
\begin{align}
N_{\mathrm{bit}}=\mathcal{O}\left(\left(N+\log_2\left(\frac{M_\mathrm{Pl}^4}{\Lambda_{\mathrm{target}}}\right)\right)2^{N/2}\right).
\end{align}

\subsubsection{Sorted-list algorithm: Grover search with a sorted list}
\label{sec:collision-algorithm}
Here, we construct a quantum algorithm for solving the vacuum-search problem through the collision problem, based on Ref.~\cite{Bernstein:2013qss}, which we refer to as the sorted-list algorithm. This algorithm can be seen as a quantum acceleration of the matching of two lists in the meet-in-the-middle algorithm.

We introduce a function-value oracle $U_Y$ that computes $\Lambda_Y(y)$:
\begin{align}
U_Y\ket{y}\ket{0}_{\mathrm{anc}}=\ket{y}\ket{\Lambda_Y(y)}_{\mathrm{anc}},
\label{eq:partial-sum-oracles}
\end{align}
where $\ket{y}$ denotes the register holding $y$ and $\ket{0}_{\mathrm{anc}}$ denotes an ancilla register initialized to $\ket{0}$, into which $\Lambda_Y(y)$ is written.
Although there are input registers that hold $V_0$ and $\Delta V_i$, we omit them from the notation for simplicity.
The oracle $U_Y$ can be constructed from a quantum adder, similar to $U_\Lambda$ in Sec.~\ref{sec:vacuum-oracle}, and corresponds to restricting the input of $U_\Lambda$ to $y$. 

We assume a quantum-accessible classical memory (QRACM, see App.~\ref{app:qracm})~\cite{Bernstein:2013qss} for the sorted-list algorithm. 
First, we compute $\Lambda_X(x)$ classically for all elements $x$ in $X$, and create a list $L$ of pairs $(x,\Lambda_X(x))$ sorted in ascending order of $\Lambda_X(x)$. This operation is the same as creating $L_X$ in step 1 of Sec.~\ref{sec:classical-mitm}. 
Next, we prepare the initial state for Grover search, which consists of a superposition over all elements of $Y$ and an ancilla register initialized to $\ket{0}$:
\begin{align}
\ket{\phi_0}=\frac{1}{\sqrt{|Y|}}\sum_{y\in Y}\ket{y}\ket{0}_{\mathrm{anc}}.
\label{eq:list-grover-initial-state}
\end{align}
The phase oracle $U_{\mathrm{Grover}}$ for Grover search is constructed in the following steps:
\begin{enumerate}
    \item Use $U_Y$ to compute $\Lambda_Y(y)$ for the bit string $y$:
    \begin{align}
    U_Y\left(\frac{1}{\sqrt{|Y|}}\sum_{y\in Y}\ket{y}\ket{0}_{\mathrm{anc}}\right)
    =\frac{1}{\sqrt{|Y|}}\sum_{y\in Y}\ket{y}\ket{\Lambda_Y(y)}_{\mathrm{anc}}.
    \end{align}    
    Note that $\Lambda_Y(y)$ is not sorted.
    
    \item Find $x\in X$ such that $0\le \Lambda_X(x) + \Lambda_Y(y) \le \Lambda_{\mathrm{target}}$ using binary search (Fig.~\ref{fig:mitm-schematic}(b)).
    In this step, we use QRACM. Let $L$ be the list sorted in ascending order, and let the $i$-th element be $(x_i,\Lambda_X(x_i))$ (i.e., $\Lambda_X(x_1)\le\Lambda_X(x_2)\le\cdots\le\Lambda_X(x_{|X|})$). An access to the list using QRACM can be expressed as
    \begin{align}
    U_{\mathrm{QRACM}}\ket{i}_{\mathrm{addr}}\ket{0}_{\mathrm{val}}=\ket{i}_{\mathrm{addr}}\ket{\Lambda_X(x_i)}_{\mathrm{val}}.
    \label{eq:list-qracm-lookup}
    \end{align}
    Here, $\ket{i}_{\mathrm{addr}}$ is the address register corresponding to the $i$-th element of $L$, and $\ket{0}_{\mathrm{val}}$ is the value register that receives the read value.
    Both registers are initialized to $\ket{0}$.
    
    Step 2 consists of repeating this access and comparison $K=\lceil\log_2|X|\rceil$ times. Let $W_k$ denote the unitary representing the $k$-th iteration. $W_k$ is composed of (i) writing the element number $i_k$ determined by the binary search, (ii) applying $U_{\mathrm{QRACM}}$ in Eq.~\eqref{eq:list-qracm-lookup}, and (iii) comparing $\Lambda_X(x_{i_k})$ in the value register with $-\Lambda_Y(y)$ to determine whether the sum is below or above zero, and writing the result to the comparison result register. The comparison results of each iteration are retained, and the address and value registers are reset to $\ket{0}$ before the next iteration. After $K$ iterations, we obtain the smallest index $i^*$ such that $\Lambda_X(x_{i^*})\ge-\Lambda_Y(y)$.\footnote{Before the binary search, we check whether $\Lambda_X(x_{|X|})\ge-\Lambda_Y(y)$. If this condition fails, no element of the list satisfies the lower bound, and we set $f(y)=0$.  
    This preliminary check is included in $U_{\mathrm{binary}}$ and does not change the asymptotic complexity.} 
    We read out $\Lambda_X(x_{i^*})$ using Eq.~\eqref{eq:list-qracm-lookup}, and check whether $\Lambda_X(x_{i^*}) + \Lambda_Y(y) \le \Lambda_{\mathrm{target}}$ holds.
    If it holds, we set $f(y)=1$, otherwise $f(y)=0$.
    We denote this operation as $C_{\mathrm{flag}}$, and then reverse all iterations. The unitary representing the entire binary search is denoted as $U_{\mathrm{binary}}$:
    \begin{align}
    &U_{\mathrm{binary}}\equiv (W_K\cdots W_1)^\dagger\,C_{\mathrm{flag}}\,(W_K\cdots W_1),\nonumber\\
    &U_{\mathrm{binary}}\ket{y}\ket{\Lambda_Y(y)}_{\mathrm{anc}}\ket{0}_{\mathrm{addr}}\ket{0}_{\mathrm{val}}\ket{0}_{\mathrm{flag}}
    =\ket{y}\ket{\Lambda_Y(y)}_{\mathrm{anc}}\ket{0}_{\mathrm{addr}}\ket{0}_{\mathrm{val}}\ket{f(y)}_{\mathrm{flag}},
    \label{eq:binary-search-unitary}\\
    &f(y)=
    \begin{cases}
    1 & \text{if}\ \exists x\in X:\ 0\le\Lambda_X(x)+\Lambda_Y(y)\le\Lambda_{\mathrm{target}}\\
    0 & \text{otherwise}
    \end{cases}.
    \nonumber
    \end{align}

    \item Perform a phase flip on $\ket{1}_{\mathrm{flag}}$, and then return the flag register to $\ket{0}_{\mathrm{flag}}$ using $U_{\mathrm{binary}}^\dagger$. This is expressed as
    \begin{align}
    U_{M'}=U_{\mathrm{binary}}^\dagger Z_{\mathrm{flag}}U_{\mathrm{binary}},\qquad
    U_{M'}\ket{y}\ket{\Lambda_Y(y)}_{\mathrm{anc}}=(-1)^{f(y)}\ket{y}\ket{\Lambda_Y(y)}_{\mathrm{anc}},
    \label{eq:list-phase-flip}
    \end{align}
    where $Z_{\mathrm{flag}}$ is the phase flip operator of the flag register.

\end{enumerate}
To summarize, the phase oracle $U_{\mathrm{Grover}}$ for Grover search is given by
\begin{align}
U_{\mathrm{Grover}}=U_Y^\dagger U_{M'} U_Y,\qquad
U_{\mathrm{Grover}}\ket{y}\ket{0}_{\mathrm{anc}}=
\begin{cases}
\ket{y}\ket{0}_{\mathrm{anc}} & \text{if} \ f(y)=0\\
-\ket{y}\ket{0}_{\mathrm{anc}} & \text{if} \ f(y)=1
\end{cases}.
\label{eq:list-grover-oracle}
\end{align}
Therefore, this oracle has the same structure as Eq.~\eqref{eq:vacuum-grover-oracle} in Sec.~\ref{sec:grover-vacuum-search}, and Grover search can be performed in the same way.
Once a $y$ with $f(y)=1$ has been found, we perform a classical
binary search in $L$ to find a corresponding $x$ and output
$\bm m=(x,y)$.

We summarize the computational complexity of the sorted-list algorithm.
Generating the list $L$ classically requires $2^p$ partial sum calculations and $\mathcal{O}(p2^p)$ comparisons for sorting, which is $\mathcal{O}(2^p)^*$ after suppressing polynomial factors for query complexity.
Grover search requires $\mathcal{O}(\sqrt{|Y|})=\mathcal{O}(2^{(N-p)/2})$ iterations.
Suppressing the polynomial cost of binary search in each iteration, the overall complexity is
\begin{align}
\mathcal{O}(2^p+2^{(N-p)/2}).
\end{align}
Choosing $p=N/3$ balances the two exponential terms, yielding
\begin{align}
\mathcal{O}(2^{N/3}).
\end{align}

We also summarize the required resources. The data used by the algorithm is divided into the list $L$ and the quantum register. The number of elements in the list $L$ is $|X|=2^p$, and each element consists of $p$ bits representing $x$ and $\mathcal{O}(\log_2(M_{\mathrm{Pl}}^4/\Lambda_{\mathrm{target}}))$ bits representing $\Lambda_X(x)$. Since $L$ is stored in classical memory, the required number of classical memory bits is
\begin{align}
N_{\mathrm{bit}}=\mathcal{O}\left(\left(p+\log_2\left(\frac{M_{\mathrm{Pl}}^4}{\Lambda_{\mathrm{target}}}\right)\right)2^p\right).
\end{align}
The quantum register consists of $N-p$ qubits for the Grover search register holding $y$, $\mathcal{O}(\log_2(M_{\mathrm{Pl}}^4/\Lambda_{\mathrm{target}}))$ qubits for the ancilla register recording $\Lambda_Y(y)$, $\log_2|X|=p$ qubits for the address register,  and $\mathcal{O}(\log_2(M_{\mathrm{Pl}}^4/\Lambda_{\mathrm{target}}))$ qubits for the value register. Therefore, the number of qubits, excluding the input registers holding $V_0$ and $\Delta V_i$, is\footnote{In this paper, we represent 1 bit of classical information using the computational basis $\ket{0},\ket{1}$ of a single qubit, and we identify the number of bits of data stored in the quantum register with the number of qubits.}
\begin{align}
N_{\mathrm{qubit}}=\mathcal{O}\left(N+\log_2\left(\frac{M_{\mathrm{Pl}}^4}{\Lambda_{\mathrm{target}}}\right)\right).
\end{align}
When the input registers holding $V_0$ and $\Delta V_i$ are included,
\begin{align}
N_{\mathrm{qubit}}=\mathcal{O}\left(N+(N-p)\log_2\left(\frac{M_{\mathrm{Pl}}^4}{\Lambda_{\mathrm{target}}}\right)\right).
\end{align}

\subsubsection{Quantum-walk algorithm: claw finding by quantum walk}
\label{sec:quantum-walk-vacuum-search}
As an alternative to the above algorithm, the quantum walk (Szegedy walk) is known to be applicable to search problems~\cite{Szegedy:2004um,Portugal:2018szw}. 
In this approach, the search space is mapped onto a graph called a Markov graph, and the existence of a solution is determined using the states represented on that graph.
For the collision problem, there is a known quantum-walk algorithm for claw finding~\cite{Tani:2007qb}.
For the window condition considered here, we use the extension
to general relations discussed immediately after Corollary~5
in Ref.~\cite{Tani:2007qb}.
Here we just summarize the result of the algorithm. 

Each node of the graph corresponds to a pair of subsets $(X',Y')$, where $X'\subseteq X$ contains $r_X$ elements and $Y'\subseteq Y$ contains $r_Y$ elements.
This algorithm implicitly uses quantum access quantum memory (QRAQM, see App.~\ref{app:qraqm}) for dealing with subsets on a node.
The query complexity of finding a collision, counting calls to the function-value oracles, is given by
\begin{align}
\mathcal{O} \left( r_X+r_Y + \sqrt{\frac{|X||Y|}{r_X}}\right).
\label{eq:claw-detection-query-complexity}
\end{align}
In our case, $|X|=2^p$ and $|Y|=2^{N-p}$, and if we choose
\begin{align}
&\frac{1}{3}N\le p\le \frac{2}{3}N,
&& r_X=r_Y=\mathcal{O}\left((|X||Y|)^{1/3}\right),
\end{align}
then the query complexity becomes
\begin{align}
\mathcal{O}\left((|X||Y|)^{1/3}\right)=\mathcal{O}(2^{N/3}).
\end{align}
The difference from the sorted-list algorithm is the freedom in choosing the partition point $p$. In the sorted-list algorithm, $p$ is fixed at $N/3$, while in the quantum-walk algorithm, any $p$ within the above range can be chosen.

Notice that the quantum-walk algorithm requires a large number of qubits to represent the Markov graph. From the dimension of the Hilbert space required to represent the graph, we have
\begin{align}
2^{N_{\mathrm{qubit}}} = \binom{|X|}{r_X} \times \binom{|Y|}{r_Y}.
\end{align}
From this equation, we can estimate
\begin{align}
N_{\mathrm{qubit}} = \mathcal{O}(r_X \log{(|X|/r_X)} + r_Y \log({|Y|/r_Y)}).
\end{align}
In the case of $r_X=r_Y=\mathcal{O}\left((|X||Y|)^{1/3}\right)$, we have
\begin{align}
    N_{\mathrm{qubit}}= \mathcal{O}\left(N 2^{N/3}\right).
\end{align}
The above estimate concerns the representation of graph vertices.
Including registers that store the contributions to vacuum energy changes the prefactor from $N$ to $N+\log_2(M_{\mathrm{Pl}}^4/\Lambda_{\mathrm{target}})$, while leaving the exponential factor $2^{N/3}$ unchanged.

\section{Comparison of classical and quantum algorithms}
\label{sec:classical-quantum-comparison}
In this section, we compare the computational complexity of the quantum vacuum-search algorithms developed in the previous section with a pseudo-polynomial time\footnote{Pseudo-polynomial time refers to an execution time that is bounded by a polynomial in the number of inputs $N$ and the magnitude of the numerical values contained in the input ($\sum_i\Delta V_i/\delta$ for our case).
The subset sum problem is a classic example of a problem that is NP-complete but has a pseudo-polynomial time algorithm.
The algorithm in Ref.~\cite{Denef:2006ad} is one such example.
} classical algorithm~\cite{Denef:2006ad} (hereafter referred to as the DD algorithm) as a classical benchmark. 
The DD algorithm determines whether there exists a bit string $\bm m$ that satisfies the vacuum energy requirement under the choice $\Delta V_i \ge 0$.\footnote{This can be done without loss of generality. For $i$ such that $\Delta V_i < 0$, replacing $m_i$ with $1 - m_i$ reverses the sign of $\Delta V_i$, with the simultaneous shift $V_0 \to V_0 + \Delta V_i$.} The computational complexity, measured by the number of iterations required to solve the problem, is
\begin{align}
\mathcal{O} \left(\frac{N \sum_i\Delta V_i}{\delta} \right),
\end{align}
where $\delta$ represents the scale of the error in $\Delta V_i$.
Given $\Lambda_{\mathrm{target}}$, we should choose $\delta$ such that $N\delta < \Lambda_{\mathrm{target}}$ to ensure that the vacuum search is meaningful.\footnote{For the comparison in this section, we exclude instances with vacuum energies within the rounding uncertainty, of order $N\delta$, of either boundary of the target window.} 
Therefore, choosing $\delta$ of order $\Lambda_{\mathrm{target}}/N$, the computational complexity of the DD algorithm becomes
\begin{align}
\mathcal{O} \left(\frac{N^2 \sum_i\Delta V_i}{\Lambda_{\mathrm{target}}}\right).
\end{align}
Since it is natural to take $\Delta V_i \sim M_{\mathrm{Pl}}^4$, the computational complexity can be expressed as
\begin{align}
\mathcal{O}\left(N^3\,\frac{M_{\mathrm{Pl}}^4}{\Lambda_{\mathrm{target}}}\right).
\end{align}

\begin{figure*}[htbp]
  \centering
  \includegraphics[width=0.8\textwidth]{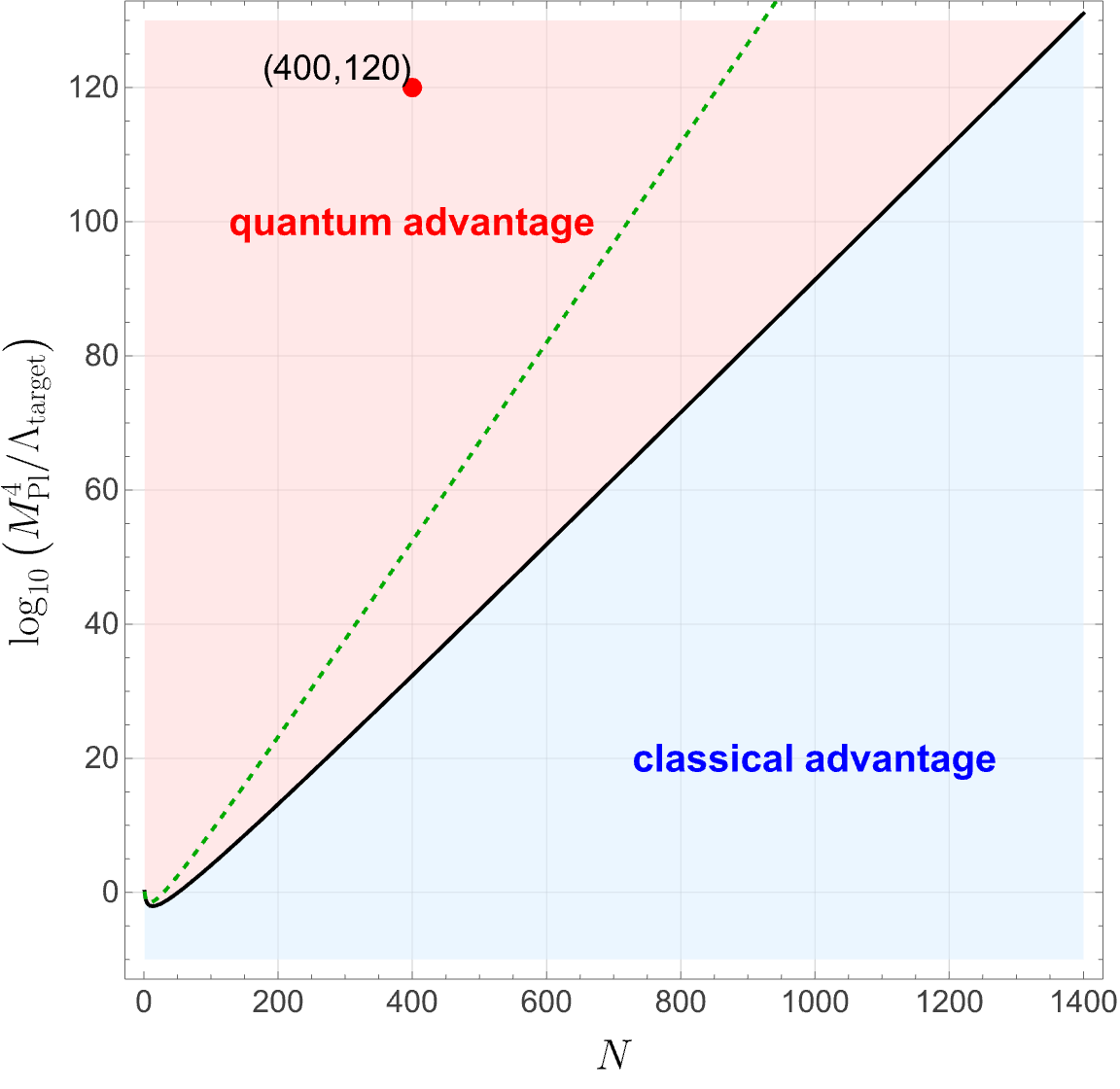}
  \caption{The comparison of the computational complexity between the DD algorithm and the quantum algorithms developed in this study. The vertical axis represents $\log_{10}(M_{\mathrm{Pl}}^4/\Lambda_{\mathrm{target}})$. The red and blue regions indicate where the quantum algorithms and the DD algorithm have lower computational complexity, respectively. The green dashed line represents the equality of the computational complexity between the DD algorithm and the plain Grover search/classical meet-in-the-middle algorithm. The red point indicates the physically motivated parameters $(N,\log_{10}(M_{\mathrm{Pl}}^4/\Lambda_{\mathrm{target}}))=(400,120)$.} 
  \label{fig:complexity-comparison}
\end{figure*}

Now we compare the DD algorithm with the quantum vacuum-search algorithms assuming that the time taken for one iteration of the DD algorithm is equal to the time taken for one execution of the oracle in the quantum algorithms.
This comparison retains the explicit factor $N^3$ in the DD estimate while suppressing polynomial prefactors in the quantum estimates.
Such prefactors do not change the displayed exponential dependence, but they can shift the crossover curves. Actual running times depend on the costs of oracle evaluation, memory access, and other operations.
Interpreting the quoted exponential factors as time scalings requires these operations, including quantum-walk data updates and marked-vertex checks, to have only polynomial overhead.
The figure therefore illustrates the stated cost model and does not predict actual running times.
The results are shown in Fig.~\ref{fig:complexity-comparison}.
The horizontal axis represents the number of fields $N$, and the vertical axis represents the magnitude of the input value $\log_{10}(M_{\mathrm{Pl}}^4/\Lambda_{\mathrm{target}})$. 
In the red region, the collision-based algorithms are more efficient, while in the blue region, the DD algorithm is more efficient.
The green dashed line indicates where the computational complexity of the DD algorithm is equal to that of the plain Grover search/classical meet-in-the-middle algorithm. 

The red point in the figure represents the physically motivated parameters where $\Lambda_{\mathrm{target}}/M_{\mathrm{Pl}}^4\sim10^{-120}$ and $N\sim400$. 
The former is motivated by the observed value of the cosmological constant.
For this benchmark, we assume that zero lies within the bulk of the vacuum-energy distribution, whose width is
$\sigma\sim\sqrt{N}M_{\mathrm{Pl}}^4$.
This is consistent with the statistical setup of Ref.~\cite{Bao:2017qmt}, where the energies of the individual minima are drawn from distributions with zero mean.
Under this assumption, the estimated number of vacua in the target window is of order $2^N\Lambda_{\mathrm{target}}/(\sqrt{N}M_{\mathrm{Pl}}^4)$.
Requiring this number to be of order unity or larger motivates $N\simeq400$ for $\Lambda_{\mathrm{target}}/M_{\mathrm{Pl}}^4\sim10^{-120}$.
Within this cost model, Fig.~\ref{fig:complexity-comparison} places this point in the region where quantum algorithms have an advantage.

Using a polynomial factor $\mathrm{poly}(N)$, the gate complexity of the quantum algorithms can be estimated as
\begin{align}
    \mathcal{O} \left(\mathrm{poly}(N)2^{N/3}\right).
\end{align}
This polynomial factor represents the gate cost of all operations per query. To keep the quantum advantage over the DD algorithm even when this factor is taken into account, we need
\begin{align}
\mathrm{poly}(N)<\frac{N^3 M_{\mathrm{Pl}}^4/\Lambda_{\mathrm{target}}}{2^{N/3}}.
\end{align}
From this inequality, the degree of $\mathrm{poly}(N)$ must satisfy
\begin{align}
    \mathrm{deg}\ \mathrm{poly}(N) <3 + \frac{\log_{10}(M_{\mathrm{Pl}}^4/\Lambda_{\mathrm{target}})}{\log_{10}N}-\frac{N \log_{10}2}{3 \log_{10}N}.
\end{align}
For the physically motivated values $\Lambda_{\mathrm{target}}/M_{\mathrm{Pl}}^4\sim10^{-120}$ and $N\sim400$ to enter the quantum advantage region, the degree of $\mathrm{poly}(N)$ must satisfy
\begin{align}
    \mathrm{deg}\ \mathrm{poly}(N) < 34.
\end{align}
Therefore, if the operations per query can be implemented at a cost lower than $\mathcal{O}(N^{34})$, a quantum advantage in gate complexity can also be realized.

\section{Conclusion and discussion}
\label{sec:conclusion}
In this paper, we have constructed quantum algorithms for vacuum search in the landscape of the ADK model. 
The most basic plain Grover search solves the ADK vacuum-search problem with a computational complexity of $\mathcal{O}(2^{N/2})$, while the collision-based algorithms (the sorted-list algorithm and the quantum-walk algorithm) achieve a computational complexity of $\mathcal{O}(2^{N/3})$.
Under the cost model of Sec.~\ref{sec:classical-quantum-comparison}, the quantum algorithms have a lower estimated computational cost than the DD algorithm for physically motivated parameters.
Similarly, compared to the classical meet-in-the-middle algorithm, the collision-based algorithms improve the exponent from $N/2$ to $N/3$.

There are several directions for future work. One is to extend the quantum algorithms constructed in this study to more realistic string theory compactification models. The main extensions are as follows:
\begin{enumerate}
    \item Generalize the parameter $m_i\in \{0,1\}$ to a parameter with a range of general integer values.
    \item Incorporate the constraints on the parameters corresponding to the tadpole condition.
\end{enumerate}
With these extensions, the quantum algorithms constructed in this study would be useful for searching for flux parameters $N^i$ that yield a small vacuum expectation value of the superpotential~\cite{Gukov:1999ya}, which is preferred in the context of the KKLT scenario~\cite{Kachru:2003aw}.
However, the coefficients of the flux parameters depend on the vacuum expectation values of the moduli, which is a difficulty in applying the algorithms in Sec.~\ref{sec:structured-vacuum-search}. 
It is necessary to identify parameter regions where the coefficients depend only weakly on the flux parameters.

Another direction for future work is to further improve the computational complexity.
The $\mathcal{O}(2^{N/3})$ scaling obtained in this study is based on the reduction to the collision problem by dividing the variables into two blocks.
On the other hand, for this type of problem, a quantum algorithm has recently been reported that achieves a computational complexity of $\mathcal{O}^*(2^{2N/7})$ with worst-case guarantees by using a different partition~\cite{Chukhin:2026qss}.
It is interesting to consider the application of this algorithm to the ADK vacuum search.

\acknowledgments
The authors would like to thank Yoshihiko Abe for carefully reading the manuscript and for helpful discussions and comments.
Y.H. was supported in part by JSPS KAKENHI Grant Nos. JP24K07035 and JP24KF0167, and by JST BOOST Program Japan Grant No. JPMJBY25E1.
OpenAI's Codex and Anthropic's Claude Code were used to assist in writing the manuscript and checking equations.

\appendix 

\section{Grover search}
\label{sec:grover-search}
We denote the set to be searched as $S$, and the set of solutions we want to find as $M$. 
We use the quantum state $\ket{s}$ to represent an element $s \in S$. Grover search~\cite{Grover:1996qsa} uses a quantum operator $U_{\mathrm{Grover}}$ (phase oracle) that determines whether a given quantum state is a solution or not:
\begin{align}
    U_{\mathrm{Grover}}\ket{s} = 
    \begin{cases}
    \ket{s} & \text{if} \ s \notin M\\
    -\ket{s}                 &  \text{if} \ s \in M 
    \end{cases}.
\label{eq:grover-phase-oracle}
\end{align}
When the number of solutions $|M|$ is known, Grover search proceeds as follows:
\begin{algorithmbox}{Algorithm: Grover search}
\begin{enumerate}
    \item Prepare the initial state
    \begin{align}
    \ket{\phi_0} = \frac{1}{\sqrt{|S|}}\sum_{s\in S} \ket{s}.
    \end{align}
    \item Apply the operator $U_{\mathrm{Grover}}$ defined in Eq.~\eqref{eq:grover-phase-oracle}.
    \item 
    Apply the diffusion operator
    \begin{align}
    2\ket{\phi_0}\bra{\phi_0}-I.
    \end{align}
    \item 
    Repeat steps 2 and 3 approximately $\frac{\pi}{4}\sqrt{\frac{|S|}{|M|}}$ times.
    \item 
    Measure the final state to read out $\ket{s}$, $s \in M$.
\end{enumerate}
\end{algorithmbox}
The query complexity of the above algorithm is
\begin{align}
\mathcal{O}\left(\sqrt{\frac{|S|}{|M|}}\right).
\label{eq:grover-query-complexity}
\end{align}
Even when the number of solutions $|M|$ is unknown, there exists an algorithm that can find a solution with the same scaling in terms of $|S|$~\cite{Boyer:1996be}. In this procedure, the number of iterations is gradually increased, and a solution is found when the optimal number of iterations is reached.
Concretely, the algorithm is as follows.
\begin{algorithmbox}{Algorithm: Grover search (unknown number of solutions)}
\begin{enumerate}
    \item Set $m=1$ and choose a fixed integer $k\ge4$.
    \item Choose an integer $j$ uniformly at random such that 
    $0 \le j < m$.
    \item Repeat steps 2 and 3 of Grover search $j$ times, and obtain the measurement result $\tilde{s}$. 
    \item Check whether $\tilde{s}$ is a solution. If it is, output $\tilde{s}$ and terminate the algorithm. If not, update $m := \min(\lambda m, \sqrt{|S|})$ and return to step 2. Here, $\lambda \in (1, 4/3)$. After reaching $m = \sqrt{|S|}$, if no solution is found for $k$ trials, output ``no solution'' and terminate the algorithm.
\end{enumerate}
\end{algorithmbox}
Here, $m$ is the upper limit of the number of Grover search iterations in one trial, and $\lambda > 1$ is the factor by which the upper limit of iterations is increased for each trial. Since $|M|$ is unknown, the optimal number of iterations is unknown. 
However, eventually $m$ will reach the level of the optimal number of iterations, and thereafter each trial will output a solution with at least a constant probability.
The condition $\lambda < 4/3$ ensures that the series of expected iterations converges when failures are repeated. Regarding the stopping condition, if $M \neq \varnothing$, after reaching $m = \sqrt{|S|}$, each trial will output a solution with at least a constant probability (evaluated as at least $1/4$ in Ref.~\cite{Boyer:1996be}), so the probability of outputting ``no solution'' when a solution exists is at most $(3/4)^k$.
For a fixed $k\ge4$, we have $(3/4)^k\le81/256<1/3$,
which ensures the success probability required in
Definition~\ref{def:adk-vacuum-search-problem}.

\section{QRAM}
\label{app:qram}
QRACM and QRAQM are two types of quantum random-access memory (QRAM) models. In this appendix, we briefly summarize the QRACM and QRAQM models.

\subsection{QRACM}
\label{app:qracm}
The sorted-list algorithm in Sec.~\ref{sec:collision-algorithm} accesses a classically prepared list $L$ for each quantum state. The device that enables this operation is called QRACM (quantum-accessible classical memory).
The QRACM was introduced in Ref.~\cite{Kuperberg:2005dhs}, and it is widely assumed in quantum algorithms for the subset sum problem, including Ref.~\cite{Bernstein:2013qss}.
In this appendix, we review the points of QRACM relevant to our algorithm.

Suppose that there are $K$ numbers of $w$-bit data $D_0,D_1,\ldots,D_{K-1}$ stored in classical memory. 
The QRACM is a device that, for an address register $\ket{i}$ consisting of $\lceil\log_2 K\rceil$ qubits and a value register $\ket{z}$ consisting of $w$ qubits, implements the operation
\begin{align}
U_{\mathrm{QRACM}}\ket{i}\ket{z}=\ket{i}\ket{z\oplus D_i},\qquad i\in\{0,1,\ldots,K-1\},\quad z\in\{0,1\}^w.
\label{eq:qracm-unitary}
\end{align}
When the address register is in a superposition of the states $\ket{i}$, from Eq.~\eqref{eq:qracm-unitary}, each component of the superposition reads out the data corresponding to its address.
It is essential to note that the data $D_i$ itself is classical and not held as a quantum state.
The assumption of QRACM means that the operation in Eq.~\eqref{eq:qracm-unitary} can be executed in $\mathcal{O}(\log K)$ time.\footnote{
As a specific implementation of QRACM, the bucket-brigade method~\cite{Giovannetti:2008qram} is known.
In this method, a binary tree is constructed where each bit of the address register switches the path, allowing access to the data at address $i$ in depth $\mathcal{O}(\log K)$.
Note that the binary tree consists of $\mathcal{O}(K)$ nodes, so the bucket-brigade QRACM requires $\mathcal{O}(K)$ physical components in total.
The evaluation of the cost including error correction has not been established~\cite{Jaques:2023qram}, so assuming that it can be executed in $\mathcal{O}(\log K)$ time is a hypothesis.
}

We assume that the QRACM is available for the sorted-list algorithm in Sec.~\ref{sec:collision-algorithm}.
In this algorithm, for each iteration of Grover search, the phase oracle $U_{\mathrm{Grover}}$ [Eq.~\eqref{eq:list-grover-oracle}] acts on the superposition of all elements of $Y$, $\sum_y\ket{y}$. The operator $U_{\mathrm{Grover}}$ computes $\Lambda_Y(y)$ for each $y$ and performs a binary search on the list $L$ to find an $x$ such that $\Lambda_X(x)$ is within the desired window.
Since the address to be read from $L$ at each stage of the binary search depends on $y$, an operation of the form in Eq.~\eqref{eq:qracm-unitary} is required.
One application of $U_{\mathrm{Grover}}$
in Eq.~\eqref{eq:list-grover-oracle}
requires $\mathcal{O}(p)$ QRACM accesses.

\subsection{QRAQM}
\label{app:qraqm}

The quantum-walk algorithm in Sec.~\ref{sec:quantum-walk-vacuum-search} accesses a memory composed of a quantum state.
The memory register itself can be in a superposition of different data sets. The device that enables this operation is called QRAQM (quantum access quantum memory).

Suppose that there are $K$ numbers of $w$-bit data $D_0,D_1,\ldots,D_{K-1}$ stored in quantum memory. 
The QRAQM is a device that, for an address register $\ket{i}$ consisting of $\lceil\log_2 K\rceil$ qubits and a value register $\ket{z}$ consisting of $w$ qubits, implements the operation
\begin{align}
U_{\mathrm{QRAQM}}\ket{i}\ket{z}\ket{D_0,D_1,\ldots,D_{K-1}}=\ket{i}\ket{z\oplus D_i}\ket{D_0,D_1,\ldots,D_{K-1}},
\label{eq:qraqm-unitary}
\\ i\in\{0,1,\ldots,K-1\},\quad z\in\{0,1\}^w. \nonumber
\end{align}
In the above equation, the third register represents the quantum memory. QRAQM allows the memory itself to be in a quantum superposition.
If the memory stores a superposition of multiple data sets, each consisting of $K$ numbers of $w$-bit data values, the $i$-th data value can be coherently accessed using the address register $\ket{i}$ while preserving the superposition of the memory.

Many quantum-walk-based algorithms assume that QRAQM is available as a data structure attached to graphs.
The claw-finding algorithm~\cite{Tani:2007qb} in Sec.~\ref{sec:quantum-walk-vacuum-search} implicitly uses QRAQM to store the information on the graph.
This structure is exploited to reduce the query complexity in the algorithm.

\bibliographystyle{PRD}
\bibliography{reference}

\end{document}